\documentclass[aps,pre,showpacs,showkeys,amsfonts,amsmath,%
twocolumn,twoside,floatfix,superscriptaddress]{revtex4}
\usepackage{amsfonts,amsmath}
\usepackage{epsfig}
\usepackage{bm}

\newcommand {\eeq}{\end{equation}}
\newcommand {\beq}{\begin{equation}}
\newcommand {\eea}{\end{eqnarray}}
\newcommand {\bea}{\begin{eqnarray}}

\begin{document}

\title{Laminar-turbulent boundary in plane Couette flow}
\author{Tobias M. Schneider}
\affiliation{Fachbereich Physik, Philipps-Universit\"at Marburg, 
D-35032 Marburg, Germany}
\author{John F. Gibson}
\affiliation{Center for Nonlinear Science, School of Physics, Georgia Institute
of Technology, Atlanta, Ga, 30332, USA}
\author{Maher Lagha}
\affiliation{Laboratoire d'Hydrodynamique (LadHyX), 
CNRS-\'Ecole Polytechnique, F-91128 Palaiseau, France}
\author{Filippo De Lillo}
\affiliation{Universit\`a degli Studi di Torino, 
Dipartimento di Fisica Generale and
Istituto Nazionale di Fisica Nucleare, sez. di Torino, Via
Giuria 1, I-10125, Torino, Italy}
\affiliation{Fachbereich Physik, Philipps-Universit\"at Marburg, 
D-35032 Marburg, Germany}
\author{Bruno Eckhardt}
\affiliation{Fachbereich Physik, Philipps-Universit\"at Marburg, 
D-35032 Marburg, Germany}
\date{\today}

\begin{abstract}
We apply the iterated edge state tracking algorithm 
to study the boundary between laminar and turbulent dynamics
in plane Couette flow at Re=400. Perturbations that are not strong enough
to become fully turbulent nor weak enough to relaminarize tend towards 
a hyperbolic coherent structure in state space, termed the edge state,
which seems to be unique up to obvious continuous shift symmetries. 
The results reported here show that in cases where a fixed point has only one unstable 
direction, as for the lower branch solution in in plane Couette flow, 
the iterated edge tracking algorithm converges to this state. 
They also show that choice of initial state is not critical, and that
essentially arbitrary initial conditions can be used to find the edge state.
\end{abstract}

\maketitle

Plane Couette flow and pipe flow belong to the class of shear
flows where turbulence occurs despite the persistent linear stability of the
laminar profile \cite{grossmann00}. 
Triggering turbulence then requires the crossing of two 
thresholds, one in Reynolds number
and one in perturbation amplitude. Guidance on the minimum
Reynolds number is offered by the appearance of exact coherent
states: once they are present, an entanglement of their stable
and unstable manifolds can provide the necessary state space
elements for chaotic, turbulent dynamics 
\cite{nagata3,nagata4,nagata5,nagata6,Cle97,waleffe98,waleffe03,%
FE_TW03,TW_Bristol,kawahara01,rrk_non,arfm,Eckh08b}. 
Since the exact coherent states
appear in saddle-node bifurcations, it is natural to associate
the upper branch (characterized by a higher kinetic energy or
a higher drag) with the turbulent dynamics and the
lower branch with the threshold in perturbation amplitude 
\cite{Toh99,itano01,waleffe01,waleffe03,Wan07a}. 
In plane Couette flow
this scenario seems to be borne out: at the point of bifurcation,
at a Reynolds number of about 127.7, the upper branch state is stable and
the lower one has only one unstable direction \cite{Cle97,waleffe03,Wan07a}. 
At slightly higher Reynolds numbers, the upper branch undergoes secondary bifurcations
which could lead to the complex state space structure usually associated
with turbulent dynamics. For the lower branch, on the other hand, there 
are no indications of further bifurcations. Since it has only
one unstable direction, its stable manifold can divide state space
such that initial conditions from one side decay more or 
less directly to the laminar profile, whereas those from the other side
show some turbulent dynamics. Such a description of the
transition along the lines of the phenomenology of saddle-node
bifurcations has been advanced by
Toh and Itano \cite{Toh99} for plane Poiseuille flow, by
Viswanath and Wang et al, \cite{Vis07,Vis07a,Wan07a} for plane Couette flow
and Kerswell and Tutty for pipe flow \cite{Ker07}.

Empirically, one may study the boundary between laminar and turbulent dynamics
by following the time evolution of flow fields 
and thereby assigning a lifetime, i.e. the time it takes for a particular 
initial condition to decay towards the laminar profile. Increasing
the amplitude of the perturbation one notes changes between regions 
with smooth variations in lifetimes (where trajectories decay rather
directly) and regions with huge fluctuations showing a sensitive dependence on initial
conditions, since neighboring initial states
can have vastly different lifetimes \cite{Sch97,Fai04,Sku06,Sch07a}. 
A point on the border between laminar and chaotic regions was said to
lie on the \emph{edge of chaos} in \cite{Sku06}. A trajectory starting
from such a point will neither decay to the laminar state nor swing up
to turbulence: it will move in regions intermediate between laminar
and turbulent motion. All points visited will lie in the edge of
chaos, and can be identified by the above search using suitable
initial conditions. What is interesting is that trajectories moving
around in this edge of chaos are dynamically attracted to a subset of
state space. This subset is invariant under the flow and attracting
for initial conditions within the edge of chaos: we call it the
\emph{edge state}. It is only a relative attractor, since it is
unstable against perturbations that lead outside the edge of chaos.

The connection between this concept and the saddle-node approach
described before is straightforward: if the boundary between laminar
and turbulent regions is formed by the stable manifold of a saddle
state, then the manifold coincides with the edge of chaos, and the
edge state is the saddle state itself.  
This is possible only if the saddle state has a single
unstable direction.  If further directions are unstable, 
the edge state will not be a fixed point but a periodic
orbit or a chaotic attractor: this seems to be the case in certain low
dimensional models and in pipe flow \cite{schneider_06,Sch07a}. 

\begin{figure}
\includegraphics[width=0.3\textwidth]{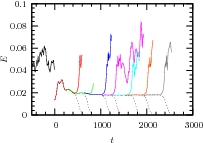} 
\includegraphics[width=0.3\textwidth]{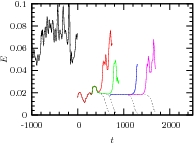} 
\includegraphics[width=0.3\textwidth]{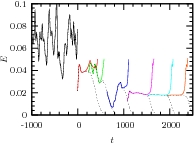} 
\caption[]{
(Color online) Time traces of the perturbation energy for three typical initial conditions, 
followed over several refinements in the edge-tracking algorithm. 
The abscissa gives the time in units of $h/U_0$ and the ordinate the energy 
per volume.
The solid lines are orbits that escape to the turbulent side, the dashed ones decay.
Time zero corresponds to the time where the edge tracking algorithm 
starts.
}
\label{e(t)}
\end{figure}

A trajectory in the edge can be bracketed by initial conditions on
the laminar and the turbulent side, i.e. initial conditions which
eventually decay or become turbulent. Technically, if we have
an initial condition ${\bf u}$ that becomes turbulent, then we study,
with ${\bf u}_L$ the laminar profile, the familiy of initial conditions
\beq
{\bf u}_\lambda={\bf u}_L+\lambda {\bf v}
\label{initial_v}
\eeq
For $\lambda=1$, this is just the continuation of the previous trajectory.
Reducing $\lambda$ the initial conditions move closer to the laminar profile
and will not become turbulent. Therefore, one can find an interval of
$\lambda$-values bounded at one end by a trajectory that becomes turbulent and
at the other by one that returns directly to the laminar profile. 
Bisecting in $\lambda$ we can focus on trajectories which live for a 
substantial time interval without becoming turbulent or decaying 
towards the laminar profile. Since the two trajectories separate exponentially,
they are followed for a finite time only, and the bisection is repeated several
times (see \cite{Sku06,Sch07a,Toh99,Vis07a} for descriptions of the
method and Fig.~\ref{e(t)} for an illustration for plane Couette flow.)
Different initial
conditions can be expected to evolve towards the same edge state,
unless there should be several, coexisting ones.

The tracking of the dynamics in the edge without the requirement of an
a priori knowledge of the hyperbolic structure offers several exciting
possibilities: first of all, it allows to study whether the dynamics in
the edge is indeed attracted to some invariant flow structures.  In
principle, as the famous examples of Julia and Mandelbrot sets
\cite{peitgen00}, and the findings for a simple model in \cite{Sku06}
show, the dynamics in the edge could be persistent and periodic or
even chaotic.  
Secondly, the convergence of the algorithm only requires the edge state to be
an attracting set, without further assumptions about its nature (fixed point,
periodic orbit or chaotic attractor).
Thirdly, the rate of separation gives
valuable dynamical information, as it limits the time interval over
which the dynamics will be close to this edge state in numerical or
experimental situations, unless further measures such as the edge
tracking algorithm are implemented.  The application to plane Couette
flow given here demonstrates the versatility of the method and allows
to connect it to the ideas about lower branch solutions advocated in
\cite{Wan07a}.  

On the numerical side, we solve the Navier-Stokes equations using a
Fourier-Chebyshev-$\tau$ scheme \cite{canuto88,Lil07a}.  The flow
domain is set to be $2$ units high, $2\pi$ units wide and $4\pi$ units
long, and periodically continued in spanwise and downstream direction.
The coordinate system is chosen such that $x$, $y$ and $z$ correspond
to the downstream, spanwise and wall-normal directions, respectively.
We use 64 Fourier modes in $x$- and 32 in the $y$-direction, and 25 Chebyshev
modes in the $z$-direction.  The equations for the wall-normal
components of velocity $w$ and vorticity $\omega_z$ were solved and
the remaining components of velocity were computed by making use of
the incompressibility condition $\nabla\cdot\mathbf{u}=0$.  The code
is fully dealiased and was verified by reproducing the linear
eigenvalue spectrum and the turbulent statistics. The Reynolds number
is based on half the channel height and half the velocity difference
between the plates. Units are chosen such that the velocity of the top
plate is $U_0=1$ and time is measured in units of 
$h/U_0$, where $h$ is half the height of the
channel. All calculations shown here are for 
a Reynolds number $Re=U_0h/\nu=400$.

\begin{figure}
\begin{tabular}{cc}
\includegraphics[width=0.20\textwidth]{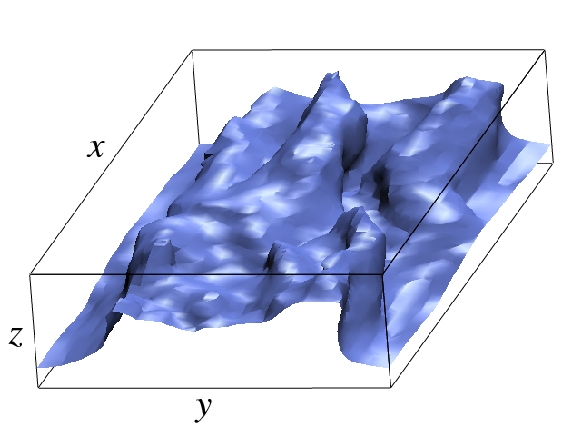} &
\includegraphics[width=0.20\textwidth]{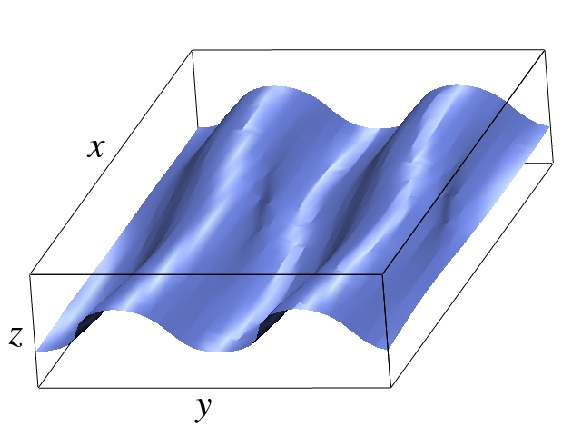} \\ 
\includegraphics[width=0.20\textwidth]{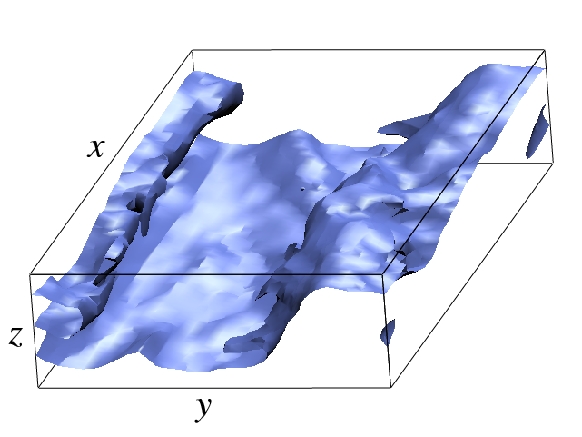} &
\includegraphics[width=0.20\textwidth]{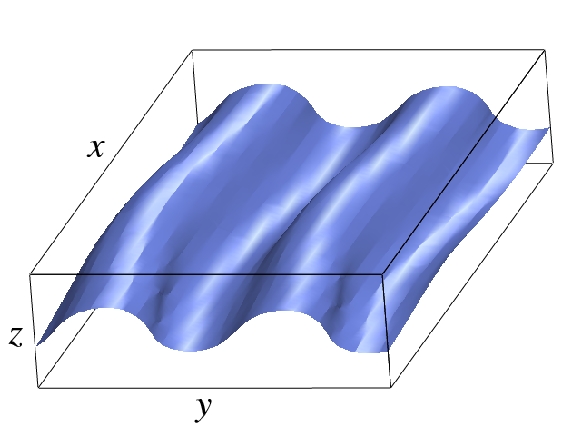} \\ 
\includegraphics[width=0.20\textwidth]{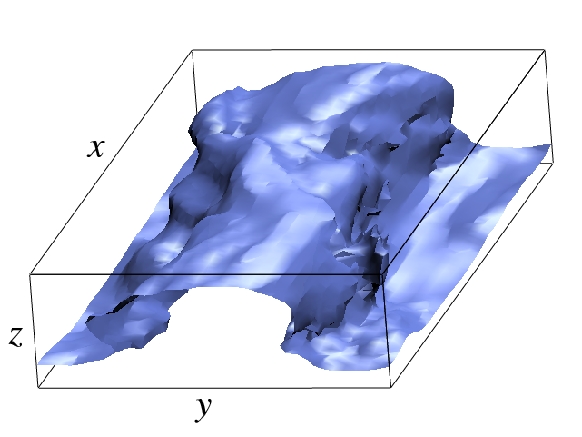} &
\includegraphics[width=0.20\textwidth]{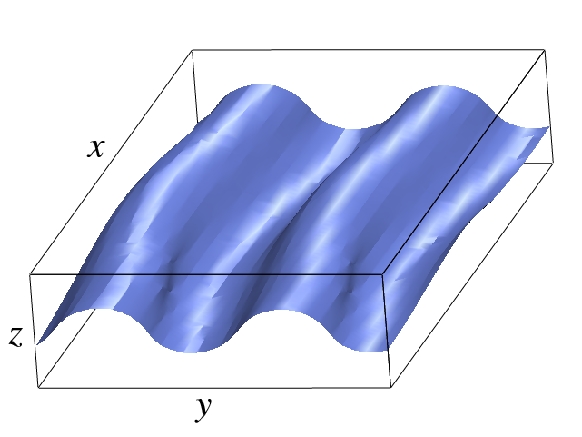} \\ 
\end{tabular}
\caption[]{(Color online) Isosurfaces for $v_x=0$ for the three trajectories
shown in Fig.~\ref{e(t)}. The left column shows the initial state
taken from the turbulent run, the right column the final state obtained
after iteration times $1568$, $1250$ and $2100$. Note the
reduced fluctuations. The final states coincide up to a shift in the
downstream direction.
}
\label{states}
\end{figure}

\begin{figure}
\includegraphics[width=0.40\textwidth]{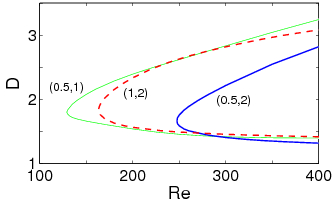} 
\caption[]{(Color online) Bifurcation diagram for three pairs of 
lower/upper-branch equilibrium solutions, with $(\alpha,\gamma)=(0.5,1)$,
$(0.5,2)$ and $(1,2)$.  The drag $D$ is the wall drag normalized
by the drag of laminar flow at the same Reynolds number. For $Re=400$, the 
$(0.5,2)$-state has the lowest drag.}
\label{wally}
\end{figure}

The time traces in Fig.~\ref{e(t)} show that independent of the
initial condition the edge tracking algorithm converges to one with
constant energy. The initial and final states shown in
Fig.~\ref{states} emphasize the very different topologies in the
initial conditions, and the coincidence in the final states, modulo
obvious translational shifts in the downstream or spanwise direction.
Wang et al \cite{Wan07a} suggested that this state should coincide
with a particular lower branch solution. Their state is similar to,
but, as pointed out by F. Waleffe (private communication),
not identical to the state found here. Introducing the
wavenumbers $\alpha$ and $\gamma$ for the structures, so that the periods 
in the downstream and spanwise direction are $2\pi/\alpha$ and $2\pi/\gamma$,
the state found by the edge tracking algorithm is $(\alpha,\gamma)=(0.5,2)$,
whereas \cite{Wan07a} suggested a state with $(\alpha,\gamma)=(0.5,1)$.
In order to explain this difference, we show in Fig.~\ref{wally}
the bifurcation diagram for three states that fit into the periodic
domain. This diagram shows the $(0.5,2)$-state to be
on the edge for $\mbox{Re}\,=400$. 
However, in view of discussion about the number of 
unstable directions, we need to verify that this state does not have
more than one unstable direction. 

\begin{figure*}
\begin{tabular}{ccc}
\includegraphics[width=0.25\textwidth]{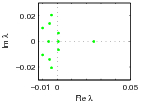} &
\includegraphics[width=0.25\textwidth]{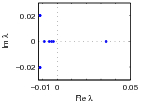} &
\includegraphics[width=0.25\textwidth]{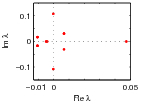} \\
\includegraphics[width=0.25\textwidth]{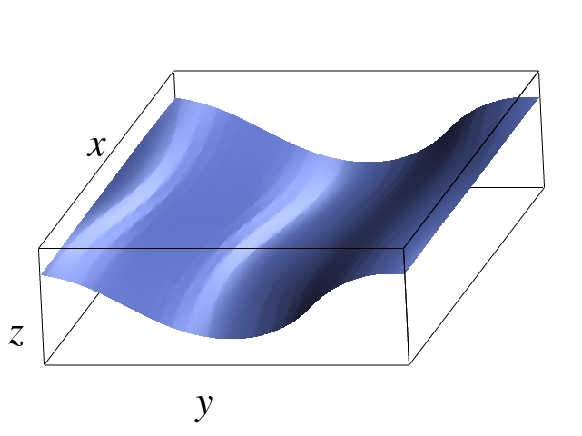} &
\includegraphics[width=0.25\textwidth]{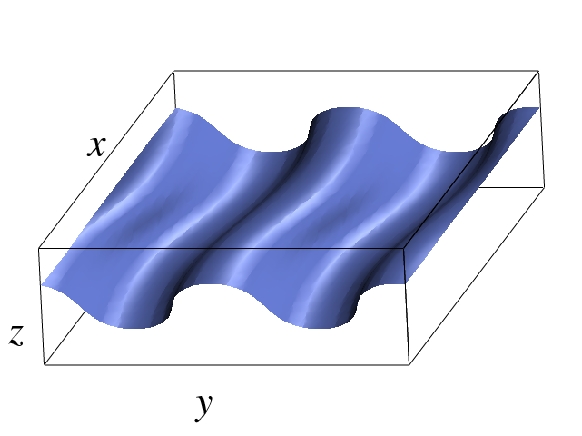} &
\includegraphics[width=0.25\textwidth]{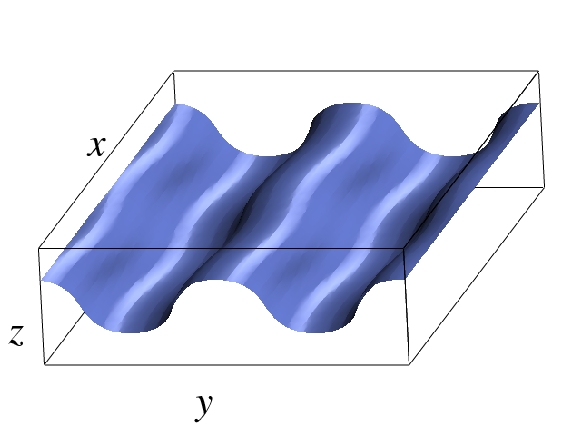} 
\end{tabular}
\caption[]{(Color online)
Eigenvalue spectra (top row) and velocity fields (bottom row)
for the states (from left to right) 
$(\alpha,\gamma)=(0.5,1)$, $(0.5,2)$ and $(1,2)$ at $Re=400$.}
\label{gibson}
\end{figure*}

Fig.~\ref{gibson} shows the $(\alpha,\gamma)=(0.5,1)$, $(0.5,2)$ and $(1,2)$
states and their eigenvalue spectra. The eigenvalues were computed with 
Arnoldi iteration on a $N_x \times N_y \times N_z = 32 \times 49 \times 64$ 
grid. Recomputing on $24 \times 35 \times 48$ 
grid confirmed their accuracy to three digits. Note that $N_y/L_y = 4N_x/L_x$,
making the spatial resolution four times finer in $y$ than in $x$. This is 
best for the problem at hand because the Fourier spectrum is far from 
isotropic. All states have two neutral eigenvalues (not shown) from the 
neutral shifts in the periodic directions downstream and spanwise. The number 
of unstable directions (where complex eigenvalues come in pairs and count as 
two directions) are $4$, $1$ and $5$ for the states $(\alpha,\gamma)=(0.5,1)$,
$(0.5,2)$ and $(1,2)$, respectively. Thus only the $(0.5,2)$ state has
only one unstable direction, consistent with the observed convergence
of the edge state tracking algorithm.

The initial conditions used here are typical in the sense that they
are taken from a turbulent velocity field. Previous studies have
focused on initial conditions with certain properties, such as
modulated downstream vortices or oblique waves. It seems that all of
them can be used to track the edge of chaos, and that all converge to
the same invariant edge state, modulo translational symmetries.

The observation that the application of the refined edge tracking method to plane
Couette flow converges to the lower branch solution, independent of initial conditions,
also has implications for the observations on pipe flow: the fact that
no steady or travelling wave invariant object is found suggests very strongly
that the invariant object is not simple, in the sense that it has more than
one unstable direction, resulting in a chaotic edge state. Obviously, if it is chaotic, there will be
simpler objects embedded, and the coherent travelling waves identified
by Pringle and Kerswell \cite{Pri07} could be first examples of such structures.

To summarize, the refined edge tracking algorithm applied here to plane
Couette flow in a parameter range where the lower branch solutions has a stable manifold
of codimension 1 confirms the expectations based on the saddle node bifurcation scenario,
in that it does converge to the lower branch solution known from the work of Nagata,
Busse and Clever \cite{Cle97}. This confirms the robustness of the algorithm and 
demonstrates that it is capable of identifying the edge state as the relative
attractor in the edge of chaos. Specifically, if the edge state has only one
unstable direction it manages to find the corresponding invariant coherent state,
since only the energy content is controlled. Conversely,
if it does not converge to a simple invariant object this
must be due to the presence of more than one unstable direction. Of course,
if the edge state is dynamically non-trivial and chaotic 
(i.e. neither a steady state nor a travelling wave) then it will contain
simpler periodic or relatively periodic structures, but they will have more
than one unstable direction so that their stable manifold has codimension
higher than one and cannot divide state space. This seems to be the case
for all coherent structures observed thus far in pipe flow.
The versatility
of the method immediately suggests applications to a variety of other
shear flows where laminar and turbulent dynamics coexist.

We thank Fabian Waleffe for his helpful comments and for his identification
of the differences in the edge states.
This work was supported in part by the Deutsche Forschungsgemeinschaft.

\end{document}